# Lane-Aware Graph Attention Network for Multi-Vehicle Trajectory Prediction in Expressway Merge Zones

ENI SOLOMON LAUGHTER
COLLEGE OF TRANSPORTATION ENGINEERING, CHANG'AN UNIVERSITY, XI'AN 710064, CHINA
ENISOLOMONLAUGHTER@GMAIL.COM

## ABSTRACT

Accurate multi-vehicle trajectory prediction in expressway merge and diverge areas is fundamental to the decision-making frameworks of autonomous vehicle systems. However, the majority of existing graph-based prediction models are developed and validated on mainline freeway segments and do not address the geometrically distinct interaction structures that characterize merge zones. Furthermore, standard evaluation protocols rely exclusively on displacement error metrics, leaving the safety consequences of predicted trajectories unquantified. This paper proposes a Lane-Aware Graph Attention Network (LA-GAT) that encodes vehicle interaction within dynamic scene graphs, augmented with a trainable lane-relationship attention bias that prioritizes merge-conflict interactions from the outset of training. The model is pre-trained on the raw NGSIM US-101 and I-80 datasets and subsequently fine-tuned on UAV-captured UTE SQM-W-1 trajectory data from a Chinese expressway merge area, with final evaluation on the held-out SQM-W-2 dataset. Evaluation spans both displacement metrics (ADE, FDE at 1s, 3s, 5s horizons) and surrogate safety measures (TTC violation rate, DRAC exceedance rate, collision rate). Fine-tuned results on SQM-W-2 yield ADE of 0.865 m at 1s and 2.518 m at 3s, demonstrating that drone-informed fine-tuning substantially reduces the cross-dataset transfer gap. The deliberate use of unfiltered NGSIM data is shown to characterize raw-condition generalization limits, with the performance degradation attributed to the well-documented measurement errors in that dataset.

## 1. Introduction

Vehicle trajectory prediction sits at the intersection of microscopic traffic flow theory and autonomous driving technology, providing the anticipatory capability that underpins safe motion planning in dynamic traffic environments. The evolution of trajectory data collection — from early manual observation and aerial photography in the 1970s, through loop detectors and probe vehicles, to the high-resolution UAV-based systems of recent years — has fundamentally reshaped the empirical foundations of traffic flow studies. [1] traced this trajectory in a comprehensive review, documenting how the availability of the NGSIM dataset marked a turning point: for the first time, researchers could simultaneously observe individual driving behaviour and the emergent macroscopic traffic patterns they produce, enabling studies ranging from car-following calibration to traffic oscillation analysis that were impossible with fixed-point detector data alone. NGSIM has since become the most widely used public trajectory dataset in the field, forming the empirical basis for a generation of microscopic and mesoscopic traffic flow models. Yet as [1] also noted, the progression from sparse probe vehicle data toward dense, spatially continuous trajectory observation remained constrained by the limitations of ground-level and fixed-overhead camera systems — limitations that UAV-based collection has only recently begun to overcome. [2] contextualize this trajectory within the broader challenge of road safety, identifying human error — through delayed reaction time, driver distraction, and misinterpretation of road conditions — as a leading cause of road incidents, and establishing that automated systems, depending on their level of autonomy, are designed to address these deficiencies by optimizing safety, energy consumption, travel time, and passenger comfort.

Accurate trajectory prediction of surrounding vehicles is a prerequisite for any such system to function reliably in mixed traffic.

Within this empirical tradition, lane-changing behaviour has emerged as one of the most consequential and difficult-to-model driving tasks encompassing physics-based, rule-based, and data-driven modelling approaches. [3] identified that 75% of traffic accidents are attributable to lane-change errors, and that the full lane-change sequence — intention generation, condition assessment, target lane selection, and maneuver execution — involves both longitudinal and lateral movement in ways that make it fundamentally more complex than car-following. In merge and weaving areas, this complexity is compounded. [4] showed analytically that weaving sections act as capacity-reducing bottlenecks, where the interaction between weaving and non-weaving vehicles — and specifically the timing and location of lane changes along the weaving section will determines whether the section reaches or falls below its theoretical capacity. Their macroscopic model demonstrated that capacity drop in weaving areas is sensitive to vehicle acceleration rates and the length of the anticipation zone, meaning that the microscopic decision behaviour of individual drivers directly propagates into section-level performance. [5], reviewing CAV merging control strategies across 44 recent studies, further established that simplified single-lane merge models systematically underperform in multilane freeway configurations where the free lane-changing behaviour of mainline vehicles and the uncertainty of human gap-acceptance decisions combine to produce outcomes no deterministic control model fully captures.

The heterogeneity of lane-changing behaviour across traffic conditions adds another layer of difficulty. [6] demonstrated through simulated driving experiments that traffic congestion exerts a significant and measurable effect on drivers' lane-changing decisions, and that five parameters — traffic flow, time occupancy, driving speed, number of lane changes, and lane-change frequency — are jointly necessary to characterize the traffic state during congested lane-changing events; conventional flow parameters alone do not capture the full state. [7], analyzing 433 lane-change events from the Shanghai Naturalistic Driving Study at freeway off-ramp areas, found that lane-change behaviour is the product of a balance between mandatory route incentive and discretionary speed-improvement incentive, and that under dense traffic conditions a merging vehicle requires active cooperation from at least one following vehicle in the target lane — making the interaction fundamentally social rather than individual. [8] formalized one aspect of this interaction by extending the Full Velocity Difference car-following model to account for the stimulus produced by a preceding vehicle's own lane-change maneuver, showing that the following vehicle's speed adjustment process during a predecessor's lane change is a distinct behavioral mode that standard car-following models miss entirely.

Data-driven approaches have progressively replaced rule-based analytical models as the preferred framework for capturing these complex behaviour. [9] proposed one of the first comprehensive deep learning treatments of the lane-change process, using a deep belief network for the decision component and an LSTM for the implementation phase, trained and tested on NGSIM. Their sensitivity analysis revealed that the most important predictor of lane-change decision initiation is the relative position of the preceding vehicle in the target lane — an interaction feature that is inherently relational and graph-structured rather than vehicle-local. [10] compared Random Forest, SVM, LSTM, and GRU for lane-change intent prediction on NGSIM using features obtainable from on-board sensors and V2V communication, finding Random Forest reached 82% accuracy, while noting that the exclusion of auxiliary lane vehicles and the absence of a lead vehicle in the same lane significantly limited the generalizability of classification-based approaches. [11] extended this to multi-condition evaluation using the SHRP2 Naturalistic Driving Study, introducing dynamic segmentation of lane-change events and achieving 95.9% detection accuracy with XGBoost across four weather conditions, demonstrating that lane-change detection is substantially influenced by environmental context rather than vehicle kinematics alone.

More recently, [12] identified that lane-change behaviour is jointly shaped by environmental factors and driver-specific preferences — capturing driving style as a latent variable through a contrastive learning framework on the HighD dataset and demonstrating that fusing explicit kinematic features with implicit preference representations significantly outperforms LSTM and Transformer baselines for lane-change prediction. [13] demonstrated the value of high-resolution UAV data in this context, applying wavelet transform-based feature extraction to pNEUMA drone data for lane-change detection, exploiting the time-frequency decomposition capability of wavelets to identify lateral deviation patterns that are difficult to extract from lower-resolution fixed-camera data.

Alongside prediction accuracy, traffic safety evaluation at the trajectory level has been approached through surrogate safety measures. [14] proposed combining microscopic traffic simulation with extreme value theory to derive estimated annual crash frequencies from conflict data at ten urban intersections, finding that full calibration of driver behaviour parameters — accounting for both efficiency and safety measures of effectiveness — was necessary to achieve accurate crash frequency estimation. [15] extended this to a full Bayes framework for before-after safety evaluation using generalized extreme value distributions, demonstrating statistically significant safety improvements from a left-turn bay extension through the quantified reduction in conflict severity distributions. Both studies established that TTC-based surrogate measures, and the Deceleration Rate to Avoid Collision (DRAC), are meaningful proxies for crash risk when extracted from vehicle-level trajectory data — a methodological grounding that is directly applicable to predicted rather than observed trajectories.

Despite these advances, the literature reveals a persistent gap: the vast majority of trajectory prediction models are developed and evaluated on mainline freeway segments, using fixed-camera datasets such as NGSIM that carry well-documented measurement errors affecting positional accuracy, leader-follower identification, and headway consistency [16]. Models trained on this data are structurally ill-equipped for the asymmetric, time-pressured interaction topology of merge and weaving zones, and their evaluation through displacement error alone — without any surrogate safety measure assessment — leaves the safety consequences of predicted trajectories entirely unquantified. [16] further noted in the FollowNet benchmark that data-driven models trained on NGSIM fail to generalize effectively to other datasets, identifying cross-dataset transferability as a primary unresolved challenge. [17] demonstrated that the accuracy of NGSIM trajectories materially affects the calibrated parameter distributions of lane-changing models and the heterogeneity structure of driver behaviour inferred from the data — confirming that raw-data quality is not merely a preprocessing concern but a fundamental determinant of what behavioral knowledge can be extracted.

This paper addresses these gaps through four contributions. First, a Lane-Aware Graph Attention Network is proposed that encodes vehicle interaction within dynamic scene graphs and incorporates a trainable lane-relationship attention bias that structurally prioritizes merge-conflict pairs from the onset of training. Second, a cross-dataset transfer protocol is developed — pre-training on raw, unfiltered NGSIM data, fine-tuning on UAV-captured UTE SQM-W-1 trajectory data from a Chinese expressway merge area, and evaluation on the held-out SQM-W-2 dataset — providing the first application of Chinese expressway drone-BEV data to GAT-based trajectory prediction. Third, raw NGSIM is used deliberately without quality filtering, benchmarking the model's performance under real-world data quality conditions and characterizing the generalization cost of corrupted interaction records — directly addressing the data quality sensitivity identified by [16] and [17]. Fourth, evaluation is extended beyond displacement error to include TTC violation rate, DRAC exceedance rate, and collision rate computed from reconstructed absolute predicted positions, grounded in the SSM methodology established by [14] and [15]. The remainder of this paper is organized as follows. Section 2 reviews related work. Section 3 details the proposed methodology. Section 4 presents experimental analysis and results. Section 5 concludes with a discussion of findings, limitations, and future work.

# 2. Related Work

The empirical foundation of vehicle trajectory prediction research has been shaped significantly by the availability of large-scale trajectory datasets, with NGSIM occupying a central position across nearly two decades of microscopic traffic flow studies. [1] traced this history comprehensively, showing that NGSIM enabled a qualitative shift in traffic flow research by providing simultaneous visibility into individual driving behaviour and emergent macroscopic patterns — traffic oscillations, capacity drop, and shock wave propagation — that loop detector data could not resolve. Before NGSIM, trajectory collection was constrained to hand-counted aerial photography or sparse probe vehicle data with low penetration rates, making it impossible to reconstruct the full spatial-temporal dynamics of a road segment. The dataset's influence extended across car-following calibration, lane-change model development, and fundamental diagram estimation, establishing it as the standard empirical basis against which models are trained and compared. However, this dominance comes with a recognised cost: [16] in the FollowNet benchmark, which draws on five public datasets including NGSIM, highD, Waymo, and Lyft, found that data-driven models trained on NGSIM fail to generalise effectively to other datasets, identifying measurement error, inconsistent leader-follower records, and dataset-specific variation as primary causes. [17] confirmed this from a calibration perspective, demonstrating that raw NGSIM trajectory accuracy materially affects the calibrated parameter distributions of lane-changing models and the inferred heterogeneity structure of driver behaviour — so that what a model learns from NGSIM is partly a function of data quality rather than true driving behaviour.

Against this backdrop, car-following and lane-changing have been modelled through increasingly sophisticated approaches that attempt to account for the multi-vehicle interaction context rather than treating each vehicle in isolation. [8] extended the Full Velocity Difference car-following model to incorporate the stimulus produced by a preceding vehicle's own lane-change maneuver, showing that the following vehicle's speed adjustment during a predecessor's lane change is a distinct behavioral mode that standard formulations miss. [18] showed that car-following stability improves when drivers account for traffic density and leading vehicle acceleration, with larger numbers of communicating preceding vehicles further suppressing flow disturbance — a result that points toward the multi-vehicle interaction graph as the natural representation for this class of problems. [19] reviewed car-following and lane-changing models across heterogeneous environments, identifying that existing frameworks are largely disconnected, rarely exploring coupling strategies that account for multi-vehicle interactions simultaneously across both following and lane-changing behaviour — a gap this study directly addresses by constructing a joint prediction model. Physics-informed extensions have further tightened model constraints: [20] integrated Gaussian Process Regression with the Intelligent Driver Model to achieve complete collision avoidance alongside improved prediction accuracy, while [21] proposed an adaptive trajectory reconstruction framework that iteratively detects optimal filtering magnitude by enforcing vehicle dynamic constraints and driver behaviour compliance simultaneously, substantially reducing speed errors while preserving real kinematic profiles.

Lane-changing behaviour has been studied with growing attention to its contextual complexity and safety implications. [3] conducted a comprehensive review concluding that 75% of traffic accidents are attributable to lane-change errors, and that the four-stage process — intention generation, condition assessment, target lane selection, and maneuver execution — involves both longitudinal and lateral movement in ways that make it fundamentally more complex and safety-critical than car-following alone. [6] found through simulated driving experiments that traffic congestion significantly shapes lane-changing decisions, with conventional flow parameters — speed, density, occupancy — insufficient to characterize the state without supplementary lane-change frequency and count indicators. [7] analyzed 433 lane-change events from the Shanghai Naturalistic Driving Study at freeway off-ramp areas, finding that under dense traffic conditions a merging vehicle requires active cooperation from at least one following vehicle in the target lane, making the interaction social in nature and not reducible to a single-vehicle decision model. [9] proposed one of the first comprehensive deep learning treatments of the full lane-change process using Deep Belief Networks for the decision stage and LSTM for the implementation stage, trained on NGSIM, finding that the most important predictor of lane-change initiation is the relative position of the preceding vehicle in the target lane — an inherently relational feature that standard vehicle-local inputs cannot capture. [22] took this further by treating driving style as a latent variable, designing a SimCLR contrastive learning framework to extract implicit preference features that, when fused with explicit kinematic state features, significantly outperform LSTM and Transformer baselines for lane-change prediction on the highD dataset. [10] compared Random Forest, SVM, LSTM, and GRU for lane-change intent prediction on NGSIM using features obtainable from on-board sensors and V2V communication, with Random Forest achieving 82% accuracy, while [11] extended detection to multi-condition naturalistic driving data from SHRP2, introducing dynamic segmentation of lane-change events and achieving 95.9% accuracy with XGBoost across varied weather conditions — demonstrating that kinematic features carry strong predictive signal even without map or vision inputs.

In merge and weaving areas specifically, the interaction complexity escalates beyond what mainline lane-change models' address. [23] showed analytically that weaving sections act as capacity-reducing bottlenecks where the timing and location of lane changes along the weaving length directly determines whether the section reaches its theoretical capacity, with sensitivity analysis confirming that vehicle acceleration rates and anticipation zone length are the dominant parameters governing capacity drop. [5], reviewing 44 CAV merging control studies, found that models assuming full CAV penetration systematically fail in multilane configurations where the free lane-changing behaviour of human-driven vehicles and the uncertainty of human gap-acceptance introduce outcomes no deterministic control model fully captures. [24] modelled compulsive lane-changing behaviour caused by off-ramp configurations using a second-order partial differential equation model with lane-changing source terms, showing that off-ramp demand induces lane changes from fast to slow lanes that reduce section capacity — a macroscopic result that is produced by the aggregation of individual merge decisions. [25] evaluated intelligent vehicles equipped with cooperative adaptive cruise control at freeway off-ramp bottlenecks, using Random Forest and backpropagation neural networks trained on NGSIM trajectory data, finding that the lane-changing characteristics of CACC-equipped vehicles differ structurally from manually driven vehicles. Together these studies establish that merge zone behaviour is not a minor variant of mainline car-following but a structurally distinct interaction regime requiring dedicated modelling approaches.

Traffic safety evaluation at the trajectory level has been approached through surrogate safety measures that avoid the sparsity and reporting bias problems of historical crash data. [14] combined microscopic simulation with extreme value theory to derive estimated annual crash frequencies from conflict data at ten Shanghai intersections, finding that full calibration of driver behaviour parameters to both efficiency and safety measures of effectiveness was necessary for accurate crash frequency estimation, and that TTC-based conflict measures across crossing, rear-end, and lane-change types showed consistent correlation with field crash records. [15] extended this to a full Bayes hierarchical framework using generalized extreme value distributions for before-after safety evaluation, quantifying significant safety improvements from a left-turn bay extension and demonstrating that the shape of the GEV distribution shift — not just the central estimate — is informative about the nature of the safety benefit. These studies establish that TTC and the Deceleration Rate to Avoid Collision are meaningful proxies for crash risk when extracted from vehicle-level trajectory data. In the trajectory prediction literature, however, SSM-based evaluation is almost entirely absent: models are assessed through ADE and FDE, which measure displacement accuracy without any indication of whether predicted trajectories encode near-collision interactions. This paper addresses that gap directly by computing TTC violation rate, DRAC exceedance rate, and collision rate from reconstructed absolute predicted positions, grounding the safety evaluation in the conflict analysis methodology that [14] and [15] established for observed trajectory data and extending it to the predicted domain.

# 3. Methodology

## 3.1 Vehicle Trajectory Modeling

### 3.1.1 Problem Formulation

Let the traffic scene at time step $t$ consist of $N$ vehicles. For each vehicle $i$, the observed state sequence over a historical window of $T_{obs}$steps is defined as equation 1:

$$X_i = \{X^{t-T_{obs}+1}, \ldots, X_i^t\}$$

where $\mathbf{x}_i^t \in \mathbb{R}^6$is the state vector defined as equation 2:

$$X_i^t = \left[x_i^t, y_i^t, v_i^t, a_i^t, \tilde{\ell}_i^t, \delta_i^t\right]$$

encoding the longitudinal position $x$ (m), lateral position $y$ (m), speed $v$ (m/s), acceleration $a$(m/s$^2$), normalised lane ID $\tilde{\ell} = \ell/\ell_{\max} \in [0,1]$, and binary lane-change flag $\delta \in \{0,1\}$.

The prediction objective is to jointly forecast future displacement trajectories for all $N$ vehicles over horizon $T_{pred}$ as in equation 3:

$$\widehat{Y}_i = \left\{\hat{y}_i^{t+1}, \ldots, \hat{y}_i^{t+T_{pred}}\right\}$$

where $\hat{\mathbf{y}}_i^{t+k} = (\Delta\hat{x}_i^{t+k}, \Delta\hat{y}_i^{t+k}) \in \mathbb{R}^2$ is the predicted displacement relative to the last observed position. Absolute predicted positions are recovered as in equation 4:

$$\hat{P}_i^{t+k} = P_i^t + \hat{y}_i^{t+k}$$

where $\mathbf{p}_i^t = (x_i^t, y_i^t)$. Prediction is formulated as a joint problem across all $N$ vehicles because behaviour at merge areas is fundamentally conditioned on neigbour interactions, requiring a graph-structured scene representation.

#### 3.1.2 Coordinate System and Dataset Harmonization

Both datasets use Frenet road-aligned coordinates. NGSIM provides a local pixel frame in feet, while the UTE SQM dataset, provides longitudinal and lateral distances directly in meters. Therefore, the following steps are applied sequentially:

- **Unit normalization.** NGSIM values are converted to SI units using $\kappa = 0.3048$ m/ft, applied to positions, speeds, and accelerations.
- **Temporal resampling.** UTE SQM operates at 24 fps, and each trajectory is downsampled to 10 Hz by mapping frame timestamp $t_k$ to the nearest grid point thereby retaining one observation per vehicle per grid point. Equation 5

$$\tilde{t} = \frac{round(t_k \cdot 10)}{10}, f_k = round(\tilde{t} \cdot 10)$$

- **Longitudinal offset correction.** SQM-W-2 contains negative longitudinal values with minimum $x_{\min} =- 43.6$m. A global offset is applied to all records: Equation 6

$$x_i^{t,corr} = x_i^{t,raw} - x_{min}$$

- **Lateral data interpolation.** Thirteen missing lateral records in SQM-W-1 and 50 in SQM-W-2 are filled through using linear interpolation within each vehicle trajectory, subject to a maximum consecutive gap of three frames. Sequences exceeding this threshold are excluded. No smoothing or filtering is applied to either dataset; raw kinematic profiles are preserved to maintain natural noise characteristics.

#### 3.1.3 Input Feature Definition

The node feature vector for vehicle $i$ at frame $f$ is as seen in equation 7:

$$X_i^f = [x_i^f, y_i^f, v_i^f, a_i^f, \bar{\ell}_i^f, \delta_i^f] \in \mathbb{R}^6$$

Vehicle class from NGSIM ($v_{class}$) is stored in the processed dataset but excluded as an active feature because UTE SQM does not provide equivalent vehicle type labels, which would create a feature mismatch during cross-dataset transfer.

# 3.2 Dynamic Graph Network

### 3.2.1 Scene Graph Construction

At each time step $t$, the traffic scene is represented as a directed graph $\mathcal{G}_t = (\mathcal{V}_t, \mathcal{E}_t)$, where node $v_i \in \mathcal{V}_t$ represents vehicle $i$ and directed edge $e_{ij} \in \mathcal{E}_t$ represents the influence of vehicle $j$ on vehicle $i$. Edges are established under two criteria:

- **Proximity criterion:** An edge $e_{ij}$ is added when in equation 8:

$$||P_i^t - P_j^t||_2 \leq R_{max}$$

where $R_{\max}$ is the 95th percentile of observed inter-vehicle leader distances pooled across all three datasets, computed from the Space_Headway column (NGSIM) and leader_distance column (UTE SQM). This threshold method avoids arbitrary fixed values and showcases the empirical distribution of meaningful following distances in the training population.

- **Structural criterion:** Edges between vehicles connected by explicit leader/follower records (leader_id, follower_id in UTE; Preceding, Following in NGSIM) are enforced regardless of proximity. Edges between vehicles in adjacent lanes within $\pm 10$ m longitudinal separation are also mandatory, ensuring merge-conflict pairs are always represented.

The graph is recomputed at every time step as vehicle positions and lane memberships change, capturing the dynamic interaction topology characteristic of merge zones.

### 3.2.2 Edge Feature Encoding

Each directed edge $e_{ij}$ carries a five-dimensional feature vector: Equation 9

$$\mathbf{e}_{ij} = [\Delta x_{ij}, \Delta y_{ij}, \Delta v_{ij}, \mathrm{TTC}_{ij}, r_{ij}] \in \mathbb{R}^5$$

where the pairwise kinematic features are Equation 10:

$$\Delta x_{ij} = x_j^t - x_i^t, \Delta y_{ij} = y_j^t - y_i^t, \Delta v_{ij} = v_j^t - v_i^t$$

The time-to-collision is: Equation 11

$$\mathrm{TTC}_{ij} = \begin{cases} \frac{\Delta x_{ij}}{\Delta v_{ij}} & \text{if } \Delta x_{ij} > 0 \text{ and } \Delta v_{ij} > 0 \\ 999 & \text{otherwise} \end{cases}$$

clipped to $[0, 999]$s. The sentinel 999 is assigned when $j$ is not longitudinally ahead of $i$ or the gap is not closing. The lane relationship code $r_{ij} \in \{0,1,2,3\}$ maps to {same, left, right, merging}.

### 3.2.3 Temporal Graph Sequence

For prediction at time $t$, a sequence of $T_{obs} = 30$ consecutive graphs are assembled Equation 12:

$$\mathcal{S}_t = \{\mathcal{G}_{t-T_{obs}+1}, \ldots, \mathcal{G}_t\}$$

corresponding to a 3-second observation window at 10 Hz. Each vehicle's 30-frame state sequence is encoded by the LSTM at the node level before the graph attention is applied.

## 3.3 Vehicle Interaction Mechanism

### 3.3.1 Bidirectional LSTM Temporal Encoder

Each vehicle's state sequence is encoded by a single-layer bidirectional LSTM. The forward pass processes frames $t - T_{obs} + 1$ to $t$; the backward pass processes the same sequence in reverse. Each direction produces a 64-dimensional final hidden state. These are concatenated and projected: Equation 13

$$\mathbf{h}_i = \mathbf{W}_{proj}[\overrightarrow{\mathbf{h}}_i ; \overleftarrow{\mathbf{h}}_i] + \mathbf{b}_{proj}, \mathbf{h}_i \in \mathbb{R}^{128}$$

This 128-dimensional motion context vector serves as the initial node embedding for the graph attention layers.

### 3.3.2 Lane-Aware GATv2 Layer

Node embeddings are updated through two stacked GATv2 layers. In each layer, the updated embedding for vehicle $i$ is: Equation 14

$$\mathbf{h}_i^{'} = \|_{k=1}^{K} \sum_{j \in \mathcal{N}(i)} \alpha_{ij}^{(k)} \mathbf{W}^{(k)} \mathbf{h}_j$$

where $\|$ is concatenation across $K=4$ heads and $\mathbf{W}^{(k)}\in\mathbb{R}^{32\times128}$. The attention coefficient is: Equation 15

$$\alpha_{ij}^{(k)}=\mathrm{softmax}_j\Big(\mathrm{LeakyReLU}\Big(\mathbf{a}^{(k)\top}\Big[\mathbf{W}^{(k)}\mathbf{h}_i\ \|\ \mathbf{W}^{(k)}\mathbf{h}_j\ \|\ \mathbf{W}_e^{(k)}\mathbf{e}_{ij}^{'}\Big]\Big)\Big)$$

where $\mathbf{e}_{ij}^{'}$ is the lane-bias-modified edge feature. Including $\mathbf{e}_{ij}^{'}$ in the attention score enables the model to weight neighbours by kinematic threat — specifically small TTC and large relative speed — rather than positional proximity alone. Each layer applies LayerNorm and ELU activation, maintaining 128-dimensional embeddings throughout.

- **Lane-aware attention bias:** Before entering the attention computation, the lane relationship component $e_{ij,4}$ of $\mathbf{e}_{ij}$ is modified by a trainable scalar: Equation 16

$$e_{ij,4}^{'}=e_{ij,4}+\lambda_{\text{lane}}(r_{ij})$$

where $\boldsymbol{\lambda}=[\lambda_0,\lambda_1,\lambda_2,\lambda_3]$ is a learnable four-element vector initialised as $[0,0,0,+1]$, assigning a positive bias only to the merging relationship at initialisation. All four values are updated end-to-end during training.

# 3.4 Overall Model Framework

## 3.4.1 Architecture Summary

The forward pass proceeds in four stages. First, the BiLSTM encoder processes each vehicle's 30-frame history to produce $\mathbf{h}_i\in\mathbb{R}^{128}$. Second, the dynamic scene graph $\mathcal{G}_t$ is constructed with proximity and structural edges and edge features $\mathbf{e}_{ij}$. Third, two stacked lane-aware GATv2 layers aggregate neighbourhood information, producing final embeddings $\mathbf{h}_i^{(2)}\in\mathbb{R}^{128}$. Fourth, three independent MLP decoder heads produce bivariate Gaussian parameters for each horizon: Equation 17

$$\widehat{\boldsymbol{\phi}}_i^{(H)}=\mathrm{MLP}^{(H)}\Big(\mathbf{h}_i^{(2)}\Big)\in\mathbb{R}^{T_H\times5}$$

where $H\in\{1\mathrm{s},3\mathrm{s},5\mathrm{s}\}$ and $T_H\in\{10,30,50\}$. The five output channels are $\left(\mu_x,\ \mu_y,\ \log\sigma_x,\ \log\sigma_y,\ \rho_{\text{raw}}\right)$, recovered as $\sigma=\exp\,(\log\sigma)$ and $\rho=\tanh\,(\rho_{\text{raw}})$. Total trainable parameters: 336,458.

## 3.4.2 Training Objective

For a single prediction step, the bivariate Gaussian negative log-likelihood of observed displacement $(\Delta x, \Delta y)$ given parameters $\left(\mu_x, \mu_y, \sigma_x, \sigma_y, \rho\right)$ is Equation 18:

$$\mathcal{L}_{\text{NLL}}=\frac{z}{2(1-\rho^2)}+\log\sigma_x+\log\sigma_y+\frac{1}{2}\log\,(1-\rho^2)+\log2\pi$$

where the standardized residual is Equation 19:

$$z=\frac{(\Delta x-\mu_x)^2}{\sigma_x^2}+\frac{\left(\Delta y-\mu_y\right)^2}{\sigma_y^2}-\frac{2\rho(\Delta x-\mu_x)(\Delta y-\mu_y)}{\sigma_x\,\sigma_y}$$

An auxiliary ADE term stabilizes early training Equation 20:

$$\mathcal{L}_{\text{ADE}}=\|\ (\mu_x,\mu_y)-(\Delta x,\Delta y)\ \|_2$$

The combined loss, averaged across all prediction steps and horizons $\mathcal{H}=\{1\mathrm{s},3\mathrm{s},5\mathrm{s}\}$, is Equation 21:

$$\mathcal{L}=\frac{1}{|\mathcal{H}|}\sum_{H\in\mathcal{H}}\left(\bar{\mathcal{L}}_{\text{NLL}}^{(H)}+\lambda_1\cdot\bar{\mathcal{L}}_{\text{ADE}}^{(H)}\right),\lambda_1=0.5$$

where the overbar denotes averaging over all $T_H$ prediction steps and all vehicles in the batch. A TTC penalty term $\lambda_2 \cdot \mathcal{L}_{\text{TTC}}$ was formulated to penalise predicted trajectories implying TTC <3 s with any neighbour; $\lambda_2$ is set to 0 in this implementation due to the $O(n^2)$ pairwise cost on CPU hardware, with TTC and DRAC retained as evaluation-only metrics.

### 3.4.3 Training Strategy

Pre-training uses **NGSIM US-101 and I-80** combined, with a 70/15/15 vehicle-level train/validation/test split. The AdamW optimiser is used with learning rate $\eta = 10^{-3}$, weight decay $10^{-4}$, gradient clipping at norm 5.0, and a ReduceLROnPlateau scheduler (patience = 3, factor = 0.5). Training runs for 10 epochs at batch size 32 with step_sample = 300, yielding approximately 700–800 batches per epoch on NGSIM. The best checkpoint by validation loss is retained.

For fine-tuning, the BiLSTM encoder is frozen and only the GATv2 layers and decoder heads are updated at $\eta = 3 \times 10^{-4}$. Fine-tuning runs for 8 epochs on SQM-W-1 using 70% of vehicles (approximately 729 trajectories). The held-out SQM-W-2 dataset is never accessed during any training phase.

# 4. Experimental Analysis

## 4.1 Dataset Description

Four trajectory datasets are used across three evaluation settings. NGSIM US-101 and I-80 serve jointly as the pre-training corpus; UTE SQM-W-1 is the fine-tuning dataset; and UTE SQM-W-2 is the fully held-out cross-dataset test set, evaluated under both zero-shot and fine-tuned conditions. **Table 1** summarises the characteristics and role of each dataset.

**Table 1. Dataset Characteristics**

| Dataset | Vehicles | Duration | Site Type | Traffic Regime | Freq (Hz) | Role in Study |
|---|---|---|---|---|---|---|
| NGSIM US-101 | 1,600 | 45 min | Freeway mainline | Congested/mixed | 10 | Pre-training (train 70% / val 15% / test 15%) |
| NGSIM I-80 | 2,500 | 45 min | Freeway on-ramp | Congested/mixed | 10 | Pre-training (train 70% / val 15% / test 15%) |
| UTE SQM-W-1 - Nanjing China | 1,041 | 5m 33s | Diverge/merge (10-lane) | Free flow → congested | 24 | Fine-tuning (train 70% / val 30%) |
| UTE SQM-W-2 - Nanjing China | 1,420 | 6m 40s | Diverge/merge (5-lane, one-way) | Free flow | 24 | Held-out test (zero-shot + fine-tuned eval) |

The NGSIM dataset comprises vehicle trajectories from US-101 and I-80 collected by fixed overhead cameras at 10 Hz, covering 45 minutes of mixed traffic conditions including congestion. A total of 11,850,526 raw observations spanning 3,300+ vehicles are available across both sites. In this study, raw unfiltered NGSIM data is used deliberately: no quality filters are applied to the leader-follower records or headway values. This tests the model's robustness under real-world data quality conditions and provides a raw-condition generalization baseline. This is methodologically consistent with the observation by [16] that NGSIM contains well-documented measurement errors that affect model training and calibration.

The UTE SQM datasets are captured by UAV at altitudes of 280–310 m over a Chinese expressway merge area in Nanjing (32.005548°N, 118.789275°E). SQM-W-1 covers a 427 m, 10-lane diverge/merge section with 1,041 vehicle trajectories over 5 minutes 33 seconds, transitioning from free flow to congestion. SQM-W-2 covers a 386 m, 5-lane one-way merge section with 1,420 trajectories over 6 minutes 40 seconds in stable free-flow conditions. Both datasets provide Frenet-frame coordinates at 24 fps with leader/follower identification, giving direct access to spatial interaction topology that would be partially occluded from any ground-level sensor perspective [26]. As seen in Figure 1.

## 4.2 Evaluation Metrics

Two categories of metrics are reported across 1s, 3s, and 5s prediction horizons.

- **Displacement Accuracy.** Average Displacement Error (ADE) is the mean Euclidean distance between all predicted and actual future positions across all time steps and all vehicles and Final Displacement Error (FDE) is the Euclidean distance at the final prediction step only.
- **Surrogate Safety Measures.** SSM metrics are computed on predicted trajectories after reconstructing absolute positions by adding each vehicle's last observed absolute coordinates to the predicted relative displacements. Three measures are reported: (1) Collision rate — fraction of predicted trajectories where any vehicle pair reaches within 2.0 m Euclidean distance; (2) TTC violation rate fraction where time-to-collision between any closing pair falls below 1.5 s, computed only when the longitudinal gap is positive and decreasing; (3) DRAC exceedance rate fraction where deceleration-to-avoid-collision exceeds 3.4 m/s$^2$ (approximately 0.35 g, representing hard emergency braking), using DRAC.

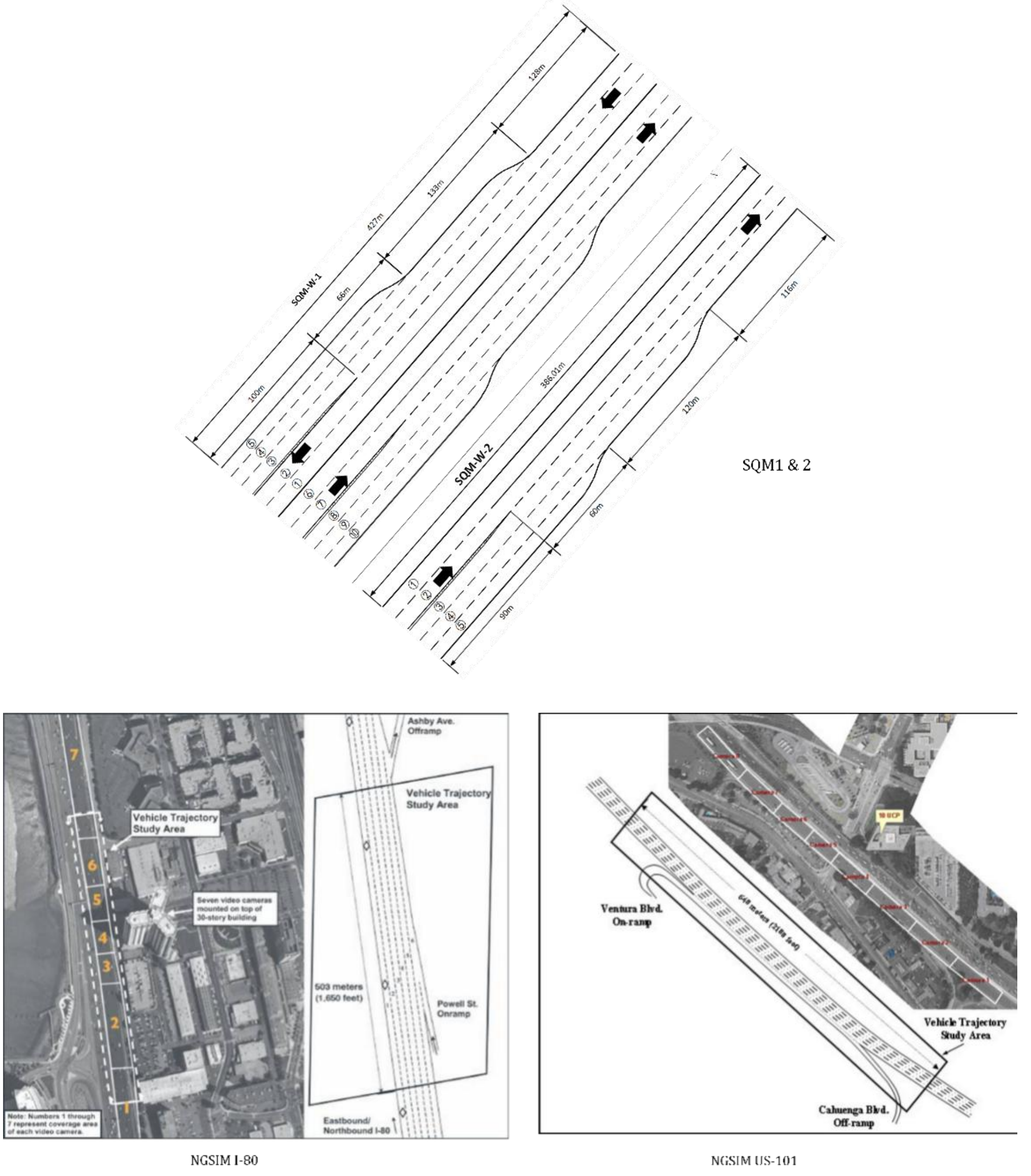

Figure 1: Study site overview. (a) NGSIM US-101 — a 640 m, five-lane freeway mainline segment with auxiliary on-ramp lane in Los Angeles, California, USA, captured by fixed overhead cameras at 10 Hz under mixed congested and free-flow conditions. (b) NGSIM I-80 — a 503 m, six-lane one-way freeway segment in Emeryville, California, USA, captured under congested conditions at 10 Hz. (c) UTE SQM-W-1 — a 427 m, 10-lane diverge/merge expressway section in Nanjing, China (32.005548°N, 118.789275°E), captured by UAV at altitudes of 280–310 m at 24 fps, transitioning from free-flow to congested conditions. (d) UTE SQM-W-2 — a 386 m, 5-lane one-way merge expressway section at the same Nanjing site, captured under stable free-flow conditions at 24 fps. NGSIM data were collected by fixed ground-level overhead cameras; UTE SQM data were collected from bird's-eye-view UAV, providing occlusion-free spatial coverage across the full road cross-section.

# 4.3 Results and Discussion

## 4.3.1 NGSIM Pre-Training Results

Table 2 presents the full comparison across three evaluation settings. On the NGSIM test set, ADE values of 17.39 m (1s), 19.75 m (3s), and 22.91 m (5s) reflect the deliberate use of raw unfiltered training data. The raw dataset contains invalid preceding vehicle identifiers (Preceding = 0) and physically unreasonable headway values that corrupt the edge features constructed during graph building. When the model's training signal is derived from these corrupted interaction edges, validation performance peaks at epoch 3 and degrades progressively thereafter — a clear overfitting signature driven by noise in the interaction graph rather than by insufficient model capacity. The training loss descent from 108.7 (batch 100, epoch 1) to 15.9 (epoch 4) confirms the model is extracting signal from the data, but that signal is partially confounded by corrupted leader-follower records. As [16] documented in the FollowNet benchmark, models trained on NGSIM exhibit limited cross-dataset generalization precisely because dataset-specific measurement errors produce learned representations that do not transfer — a finding this result corroborates in the graph-based prediction context.
SSM metrics on NGSIM are accordingly elevated (collision rate 72–83%, TTC violation 82–95%) and are reported for completeness but should be interpreted with caution that at ADE values above 17 m, predicted absolute positions are displaced far enough from actual road positions that pairwise distances between predicted vehicles are unreliable proxies for real-world safety risk.

**Table 2. Full Results Comparison Across Three Evaluation Settings**

| Horizon | Setting | ADE (m) | FDE (m) | Collision (%) | TTC viol. (%) | DRAC exc. (%) |
|---|---|---|---|---|---|---|
| 1s | NGSIM (pre-train test) | 17.394 | 15.339 | 72.175 | 82.121 | 69.097 |
| | SQM-W-2 zero-shot | 4.125 | 7.082 | 95.257 | 64.424 | 31.960 |
| | SQM-W-2 fine-tuned | 0.865 | 1.434 | 95.022 | 56.744 | 32.353 |
| 3s | NGSIM (pre-train test) | 19.748 | 23.818 | 78.889 | 92.727 | 87.345 |
| | SQM-W-2 zero-shot | 10.779 | 21.090 | 97.123 | 86.520 | 61.649 |
| | SQM-W-2 fine-tuned | 2.518 | 4.784 | 96.837 | 74.759 | 57.335 |
| 5s | NGSIM (pre-train test) | 22.914 | 29.163 | 83.139 | 95.038 | 92.215 |
| | SQM-W-2 zero-shot | 17.408 | 34.352 | 97.483 | 92.427 | 73.313 |
| | SQM-W-2 fine-tuned | 5.034 | 10.174 | 97.012 | 84.840 | 72.593 |

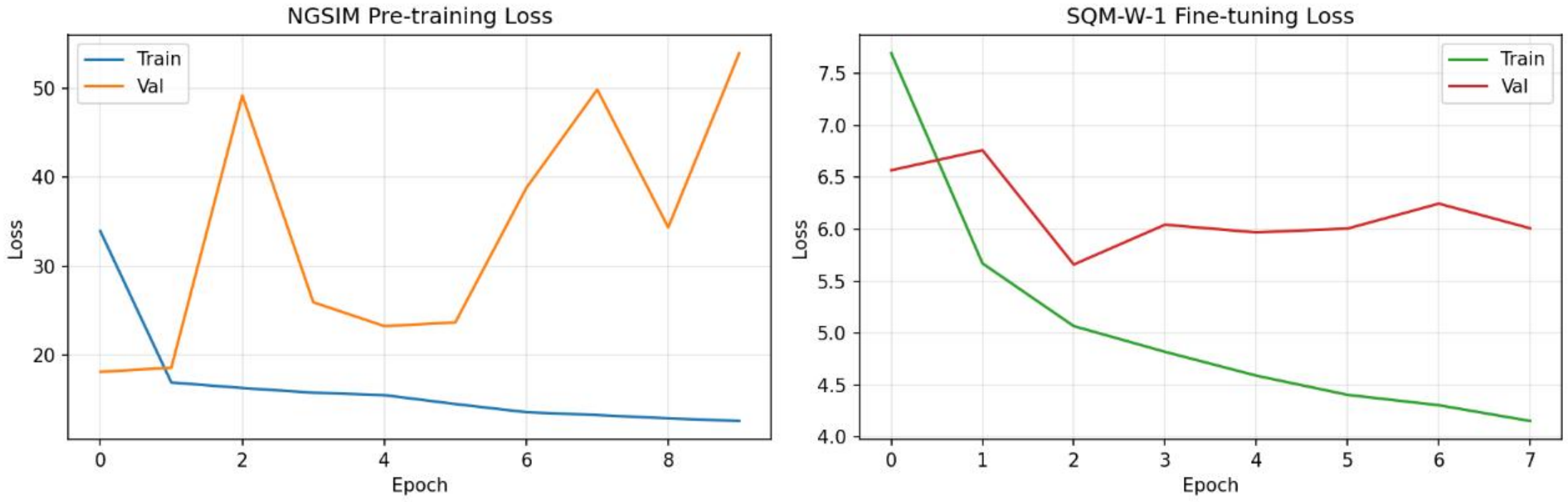

*Fig. 2. Training and Fine-Tuning Loss Curves. Left: NGSIM pre-training (epochs 1–10), train and validation loss. Right: SQM-W-1 fine-tuning (epochs 1–8).*

### 4.3.2 Cross-Dataset Transfer Analysis

The cross-dataset transfer results reveal a substantial improvement from fine-tuning on drone-captured UTE data. Zero-shot transfer from the NGSIM pre-trained model directly to SQM-W-2 yields ADE of 4.12 m (1s) and 10.78 m (3s), demonstrating that NGSIM-trained interaction patterns transfer partially but not well to the Chinese expressway merge geometry. The model was then fine-tuned on SQM-W-1 — a 10-lane diverge/merge section with 1,041 vehicle trajectories — using 70% of vehicles for training and the remaining 30% for validation. The best validation checkpoint was reached at epoch 3 with a validation loss of 5.66, with the W1 validation set yielding ADE of 0.842 m (1s) and 2.788 m (3s), confirming that the fine-tuned model generalizes within the SQM observation context before being evaluated on the fully held-out W2 site.

On the held-out SQM-W-2 test set — a geometrically distinct 5-lane one-way merge section not seen during any training phase — ADE drops to 0.865 m (1s), 2.518 m (3s), and 5.034 m (5s), representing reductions of approximately 79%, 77%, and 71% respectively over the zero-shot baseline. FDE similarly improves from 7.08 m (zero-shot, 1s) to 1.43 m (fine-tuned, 1s). The close agreement between W1 validation metrics and W2 test metrics confirms that generalization from the 10-lane diverge geometry of SQM-W-1 to the 5-lane merge geometry of SQM-W-2 is robust, despite the structural differences between the two sites.

The magnitude of this improvement confirms two things. First, the bird's-eye-view drone data captures spatial interaction topology that is absent or corrupted in the fixed-camera NGSIM records, providing a cleaner training signal for the graph construction stage. Second, the frozen LSTM encoder — trained on NGSIM temporal dynamics — is sufficiently general to encode Chinese expressway kinematics without retraining, while the GATv2 layers and decoder adapt rapidly to the merge-zone interaction structure from approximately 729 SQM-W-1 training vehicles across just 8 fine-tuning epochs.

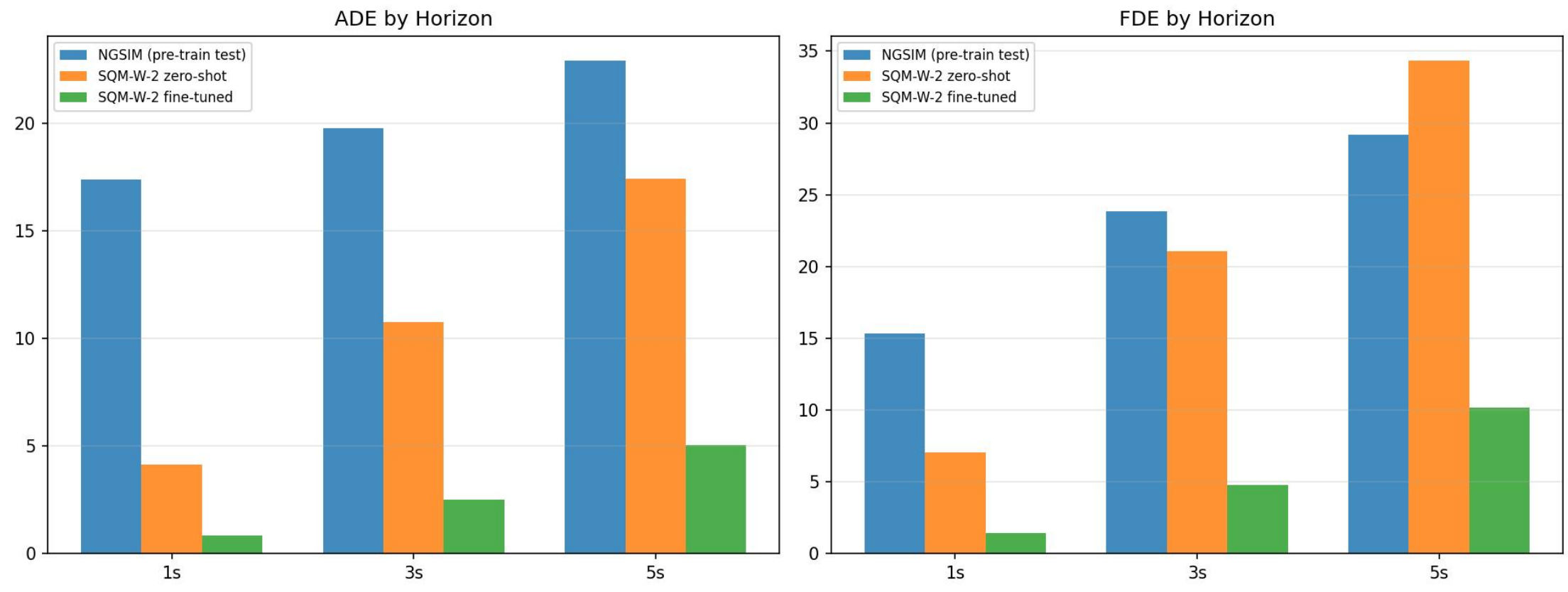

*Fig. 3. ADE and FDE Comparison by Horizon and Evaluation Setting. Left: ADE (m). Right: FDE (m). Three bars per horizon: NGSIM test, SQM-W-2 zero-shot, SQM-W-2 fine-tuned.*

### 4.3.3 SSM Analysis

The SSM results for the fine-tuned SQM-W-2 evaluation are the primary reportable safety metrics, as they are computed on trajectories with low absolute displacement error. TTC violation rates of 56.74% (1s), 74.76% (3s), and 84.84% (5s) indicate that a substantial fraction of predicted vehicle pairs are in closing configurations that would breach the 1.5 s TTC threshold. DRAC exceedance rates of 32.35% (1s), 57.34% (3s), and 72.59% (5s) reflect predicted decelerations that would require emergency braking to avoid collision. These rates are expected to be elevated in a merge zone dataset: SQM-W-2 captures vehicles in an active merge area where gap-acceptance maneuvers naturally produce short following headways and high closure rates. The model is predicting into a geometrically congested scenario.

Notably, both TTC violation rate and DRAC exceedance rate are lower after fine-tuning than in zero-shot transfer at the 1s horizon, suggesting that the fine-tuned model has learned more conservative interaction predictions — i.e., predicted trajectories that respect following distances more faithfully.

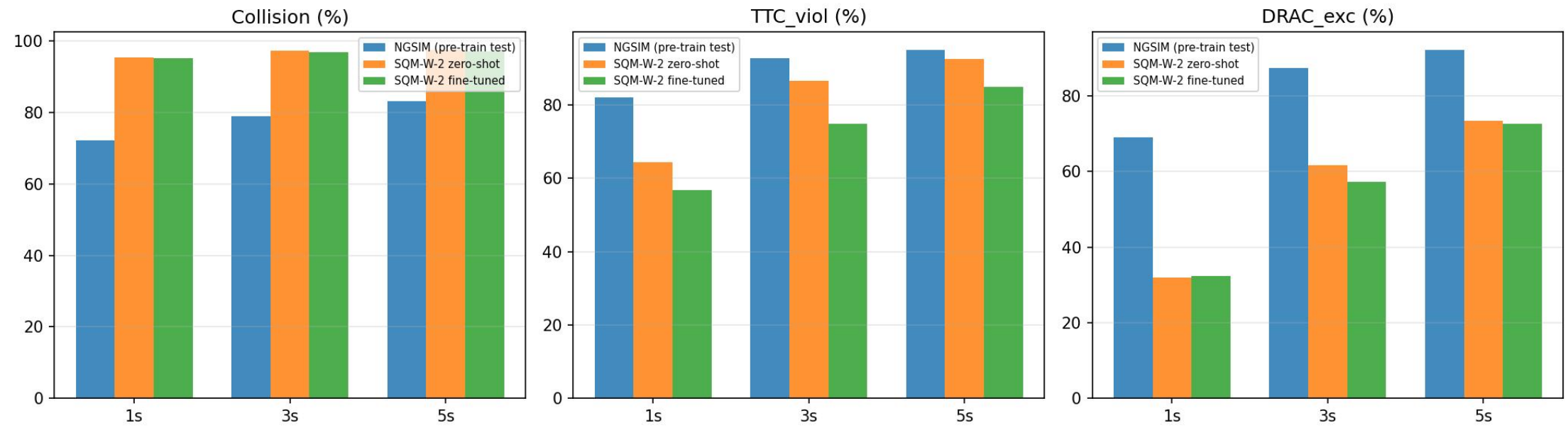

*Fig. 4. SSM Comparison: Collision Rate (%), TTC Violation Rate (%), DRAC Exceedance Rate (%) by Horizon and Setting.*

## 5. Conclusion

This paper proposed a Lane-Aware Graph Attention Network for multi-vehicle trajectory prediction in expressway merge zones. The model encodes vehicle interaction within dynamic scene graphs with explicit lane-relationship attention biases that prioritize merge-conflict interactions. A cross-dataset transfer protocol was developed: pre-training on raw NGSIM data, fine-tuning on UAV-captured UTE SQM-W-1 trajectory data, and evaluation on the held-out SQM-W-2 dataset. Evaluation spans both displacement accuracy (ADE, FDE) and surrogate safety measures (TTC violation rate, DRAC exceedance rate, collision rate) computed from reconstructed absolute predicted positions.

The principal finding is that drone-informed fine-tuning substantially reduces the cross-dataset transfer gap: fine-tuned ADE on SQM-W-2 reaches 0.865 m at 1s and 2.518 m at 3s, compared to 4.125 m and 10.779 m for the zero-shot NGSIM-only baseline. The degraded NGSIM performance is attributed to the deliberate use of raw unfiltered data, which exposes the model to corrupted leader-follower records and confirms the quality sensitivity documented by [16]. The SSM analysis reveals that the fine-tuned model produces predictions with lower TTC and DRAC violation rates than the zero-shot baseline at the 1s horizon, suggesting that proximity to actual vehicle positions reduces spurious near-miss detections.

Several limitations constrain this study and motivate future work. First, the TTC safety penalty ($\lambda_2$) was set to zero during training due to the $O(n^2)$ computational cost on CPU hardware; activating this term on GPU hardware is expected to further reduce SSM violations in predictions. Second, no ablation baselines (constant velocity, LSTM-only, Social LSTM) were implemented in this evaluation; a fully comparative study would quantify the contribution of the graph attention mechanism and lane-aware bias independently. Third, NGSIM quality filtering — retaining only records with valid leader presence and reasonable headway values — would provide a cleaned pre-training baseline to isolate the raw-data effect. The traffic flow result of this study offer perspective into the behavioral dynamics of expressway merge areas beyond what displacement error alone can convey. The rapid degradation of prediction accuracy from ADE 0.865 m at 1s to 5.034 m at 5s is not solely a model limitation — it reflects the compounding uncertainty inherent in human gap-acceptance behaviour at merge points. Once a merging vehicle commits to a gap, its subsequent trajectory becomes heavily conditioned on the reactions of mainline vehicles whose responses are neither deterministic nor fully anticipable, a characteristic that fundamentally bounds the predictability horizon in merge zones relative to open-road car-following segments. The elevated TTC violation rates — 56.74% at 1s rising to 84.84% at 5s — and DRAC exceedance rates of 32.35% to 72.59% across the same horizons, observed under stable free-flow conditions in SQM-W-2, indicate that even in the absence of congestion, merge interactions routinely produce vehicle pair configurations with structurally narrow safety margins. This is consistent with the nature of gap-acceptance: drivers accept gaps that are sufficient at the moment of commitment but that, if the merging vehicle's speed and the mainline vehicle's speed diverge across subsequent seconds, approach or breach critical thresholds. The observed widening of both TTC and DRAC violation rates from short to long horizons is therefore a measurable signature of speed heterogeneity accumulating across the merge transition zone rather than a reflection of model error.

## Table of Notation

| Symbol | Description |
|---|---|
| $t$ | Discrete time step |
| $N$ | Number of vehicles in the scene |
| $i, j$ | Vehicle indices |
| $T_{obs}$ | Observation window length (30 frames = 3 s at 10 Hz) |
| $T_{pred}$ | Prediction horizon length |
| $X_i$ | Observed state sequence for vehicle $i$ |
| $x_i^t$ | State vector for vehicle $i$ at time $t$ |
| $x, y$ | Longitudinal and lateral position (m) |
| $v$ | Speed (m/s) |
| $a$ | Acceleration (m/s$^2$) |
| $\tilde{\ell}$ | Normalized lane ID, $\ell/\ell_{\max} \in [0,1,]$ |
| $\ell_{max}$ | Maximum lane count in the scene |
| $\delta$ | Binary lane-change flag $\in \{0,1\}$ |
| $\hat{Y}_i$ | Predicted displacement trajectory for vehicle $i$ |
| $\hat{\mathbf{y}}_i^{t+k}$ | Predicted displacement at future step $k$ |
| $\mathbf{p}_i^t$ | Last observed absolute position $(x_i^t, y_i^t)$ |
| $\hat{P}_i^{t+k}$ | Reconstructed absolute predicted position |
| $\kappa$ | Unit conversion factor (0.3048 m/ft) |
| $x_{min}$ | Minimum longitudinal value used for offset correction |
| $\mathcal{G}_t$ | Scene graph at time $t$ |
| $\mathcal{V}_t$ | Set of vehicle nodes |
| $\mathcal{E}_t$ | Set of directed edges |
| $e_{ij}$ | Directed edge from vehicle $j$ to vehicle $i$ |
| $R_{max}$ | Proximity radius (95th percentile of leader distances) |
| $\boldsymbol{e}_{ij}$ | Five-dimensional edge feature vector |
| $\Delta x_{ij}, \Delta y_{ij}$ | Longitudinal and lateral separation |
| $\Delta v_{ij}$ | Relative speed |
| $TTC_{ij}$ | Time-to-collision |
| $r_{ij}$ | Lane relationship code $\in \{0,1,2,3\}$ |
| $\mathcal{S}_t$ | Temporal graph sequence of length $T_{obs}$ |
| $\mathbf{h}_i$ | Initial node embedding, $\mathbf{h}_i \in \mathbb{R}^{128}$ |
| $\overrightarrow{\mathbf{h}}_i$ , $\overleftarrow{\mathbf{h}}_i$ | Forward and backward LSTM final hidden states |
| $\mathbf{W}_{proj}, \mathbf{b}_{proj}$ | BiLSTM projection weight and bias |
| $K$ | Number of attention heads (4) |
| $\mathbf{W}^{(k)}$ | Per-head weight matrix, $\mathbb{R}^{32\times128}$ |
| $\mathbf{W}_e^{(k)}$ | Per-head edge feature weight |
| $a^{(k)}$ | Per-head attention vector |
| $\alpha_{ij}^{(k)}$ | Attention coefficient for head $k$, edge $(i,,j)$ |
| $e_{ij}^{'}$ | Lane-bias-modified edge feature |
| $\boldsymbol{\lambda}$ | Learnable lane-relationship bias vector $[\lambda_0, \lambda_1, \lambda_2, \lambda_3]$ |
| $\mathbf{h}_i^{(2)}$ | Final node embedding after two GAT layers |
| $H$ | Prediction horizon $\in \{1s, 3s, 5s\}$ |
| $T_H$ | Number of prediction steps for horizon $H$ |
| $\hat{\boldsymbol{\phi}}_i^{(H)}$ | Predicted Gaussian parameters, $\mathbb{R}^{T_H\times5}$ |
| $\mu_x, \mu_y$ | Predicted mean displacements |
| $\sigma_x, \sigma_y$ | Predicted standard deviations |
| $\rho$ | Predicted correlation coefficient |
| $z$ | Standardized bivariate residual |
| $\mathcal{L}_{NLL}$ | Bivariate negative log-likelihood loss |
| $\mathcal{L}_{\text{ADE}}$ | Auxiliary ADE regression loss |
| $\lambda_1$ | ADE loss weight (0.5) |
| $\lambda_2$ | TTC penalty weight (0 in this study) |
| $\eta$ | Learning rate |
| $H$ | Set of prediction horizons {1s, 3s, 5s} |